\documentclass[a4paper,twocolumn]{article}

\usepackage{amsthm,amsmath}
\usepackage{amssymb}
\usepackage{amsfonts}
\usepackage{url} 
\usepackage{graphicx}
\usepackage{cite}
\usepackage[affil-it]{authblk} 
\usepackage{mathptmx} 
\usepackage{flushend}

\newtheorem{theorem}{Theorem}
\newtheorem{remark}{Remark}\theoremstyle{remark}
\newtheorem{definition}{Definition}\theoremstyle{definition}

\begin{document}
	
	\title{Observability and Structural Identifiability of Nonlinear Biological Systems\footnote{This article has been accepted for publication in the special issue ``Computational Methods for Identification and Modelling of Complex Biological Systems'' of Complexity}}
	
	\author{Alejandro F. Villaverde}
	\affil{Bioprocess Engineering Group, IIM-CSIC, Vigo 36208, Galicia, Spain\\afvillaverde@iim.csic.es}
	
	\date{\today}
	
	\maketitle


\begin{abstract}	
Observability is a modelling property that describes the possibility of inferring the internal state of a system from observations of its output. A related property, structural identifiability, refers to the theoretical possibility of determining the parameter values from the output. In fact, structural identifiability becomes a particular case of observability if the parameters are considered as constant state variables. It is possible to simultaneously analyse the observability and structural identifiability of a model using the conceptual tools of differential geometry. Many complex biological processes can be described by systems of nonlinear ordinary differential equations, and can therefore be analysed with this approach. The purpose of this review article is threefold: (I) to serve as a tutorial on observability and structural identifiability of nonlinear systems, using the  differential geometry approach for their analysis; (II) to review recent advances in the field; and (III) to identify open problems and suggest new avenues for research in this area.
\end{abstract}

\section{Introduction}\label{sec:intro}

A model is observable if it is theoretically possible to infer its internal state by observing its output. 
Model parameters can be considered as constant state variables. The particular case of parameter observability is called structural identifiability. 
Both concepts are \textit{structural} in the sense that they depend only on the model equations, that is, they are completely determined by the system dynamics and output definition. They are not affected by limitations related to the frequency or accuracy of the experimental measurements, in contrast to the related concept of \textit{practical} identifiability or estimability. 

The concept of observability was introduced by Kalman in 1960 for linear time-invariant systems\cite{kalman1960contributions,kalman1960general}. Conditions for checking observability of nonlinear systems were soon developed by several authors \cite{kostyukovskii1968simple,griffith1971observability,sussmann1972controllability,kou1973observability,hermann1977nonlinear}.
At the same time, the interest in parametric identifiability was growing among researchers using biological models, especially in biomedical applications. As a result, the concept of structural identifiability was introduced in 1970, when Bellman and {\AA}str\"om coined the term and presented the Laplace transform method for its study in the context of (linear) compartmental models \cite{bellman1970structural}. 

Both concepts, observability and structural identifiability, are applicable to dynamic models of any kind: electrical, chemical, mechanical, biological, etc. Observability analysis, as well as the related question of observer design, has been and continues to be frequently investigated by systems and control theorists. In turn, researchers working in biological modelling (e.g. in mathematical biology and, more recently, in the systems biology community) have more often addressed structural identifiability issues. This is due to the fact that biological applications typically have more experimental limitations than engineering ones in terms of which measurements are feasible, making parameter identification a more challenging problem and calling for a deeper study of parametric identifiability issues and methods.  

Observability and structural identifiability play a central role in system identification. There are a number of classic books on the subject, such as the ones by Walter and Pronzato \cite{walter1997identification} and Ljung \cite{ljung1999system}. In the context of biological modelling a very complete and recent reference is the book by DiStefano \cite{distefano2015dynamic}, which covers thoroughly the topic of identifiability, both from structural and practical points of view. The interested reader is also referred to \cite{villaverde2016identifiability}, which reviews the different types of identifiability and related concepts, and to \cite{miao2011identifiability,Chis11a}, which deal specifically with structural identifiability. In a different context, Chatzis and coworkers have reviewed the observability and structural identifiability of nonlinear mechanical systems \cite{chatzis2015observability}.

The present paper reviews observability and structural identifiability concepts and tools, with the aim of facilitating their application to biological models. Instead of attempting to discuss all the existing methodologies, it focuses on methods that adopt a differential geometry approach \cite{isidori1995nonlinear,vidyasagar1993nonlinear,sontag1982mathematical}.
These properties may also be analysed with other symbolic approaches, such as power series \cite{Pohjanpalo1978,walter1982global,Chis11b}, differential algebra \cite{diop1991nonlinearCDC,ljung1994global,audoly2001global,meshkat2009algorithm,hong2018global}, or others \cite{vajda1989similarity,denis2001some,xia2003identifiability}, to name just a few, as well as with semi-numerical \cite{sedoglavic2002probabilistic,anguelova2012efficient} or numerical approaches \cite{stigter2015fast,raue2009structural}. 
A comparison or discussion of the aforementioned methods is out of the scope of the present paper; the interested reader is again referred to \cite{miao2011identifiability,Chis11a,raue2014comparison,villaverde2016identifiability}.

This manuscript begins by motivating the study in Section \ref{sec:motivation}, illustrating the possible consequences of unobservability and unidentifiability.
In Section \ref{sec:struct} these concepts are analysed with the differential geometry approach, which provides a unified view of observability and structural identifiability and can be applied to a very general class of nonlinear systems. Section \ref{sec:recent} reports recent developments in this area, and Section \ref{sec:conclusions} concludes by suggesting some open problems as possible research directions.

\section{Motivation: implications of unobservability and unidentifiability in biological models}\label{sec:motivation}

The importance of structural identifiability analysis has been recently stressed in different areas of biological modelling, such as animal science \cite{munoz2018or}, pharmacodynamics \cite{janzen2016parameter}, epidemiology \cite{tuncer2018structural}, environmental modelling \cite{stigter2017assessing}, physiology \cite{middendorf2016structural}, neuroscience \cite{walch2016parameter}, oncology \cite{saccomani2018union}, and many more.
On the other hand, assessing observability and structural identifiability can be difficult even for relatively small systems, and becomes increasingly complicated as the model complexity increases. Furthermore, the theoretical foundations of the analyses have some aspects that are not fully studied yet. These reasons help explain why some modellers are reluctant to analyse these properties of their models \cite{distefano2015dynamic}, which might be understandable taking into account that even the need of determining parameter values has been questioned in the context of biological modelling \cite{gutenkunst2007universally}. However, such analysis is worth the effort, since lack of identifiability and/or observability can compromise the ability of a model to provide biological insight \cite{janzen2016parameter,villaverde2017dynamical,eisenberg2017confidence,procopio2017model,munoz2018or}.
For example, one of the possible purposes of a model is for inferring the values of certain parameters of interest; in such case, identifiability is obviously desirable \textit{per se}. Alternatively, the main purpose of the model may be to predict the dynamic behaviour of unmeasured states; in this case one is more interested in state observability than in parameter identifiability (although issues with the latter property may compromise the former). 

\begin{figure*}[t!]
	\centering
	\includegraphics[width=1.0\linewidth]{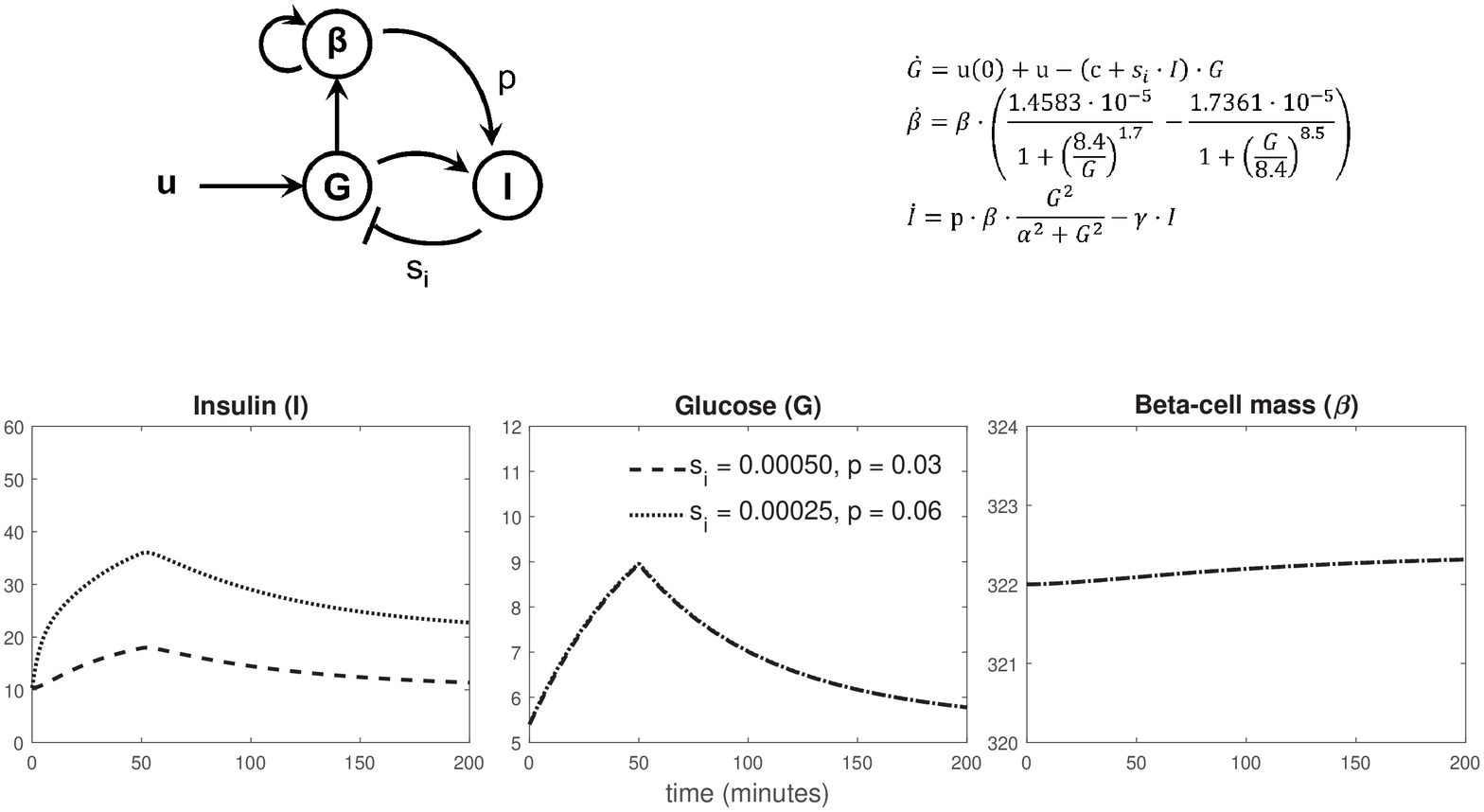}
	\caption{Illustration of observability and structural identifiability issues. Top: diagram and equations of the `$\beta$IG model' of the glucose-insulin system \cite{karin}. If glucose concentration (G) and $\beta$-cell mass are measured, the parameters $p$ and $s_i$ are structurally unidentifiable: the bottom plots show that different combinations of $p$ and $s_i$ values yield identical curves of G and $\beta$, so it is not possible to distinguish between them as long as the product $p\cdot s_i$, which is structurally identifiable, remains constant.
	Likewise, in this case insulin concentration (I) is an unobservable state: it is not possible to determine which of the two time courses of I shown in the lower left plot is the true one.}
	\label{fig1}
\end{figure*}

As an example, consider the model of a possible glucose homeostasis mechanism depicted in Figure \ref{fig1}, which was presented in \cite{karin} and analysed in \cite{villaverde2017dynamical}.
This so-called $\beta$IG model describes the regulation of plasma glucose concentration (G) by means of insulin (I), which is secreted by pancreatic $\beta$ cells. The model consists of three state variables ($\beta$,I,G) whose time courses are defined by nonlinear ordinary differential equations (ODEs) with five parameters ($c,s_i,p,\alpha,\gamma$). For the sake of the exercise, let us assume that glucose and $\beta$-cell mass are the measured outputs. In this case, if the model parameters are unknown, $p$ and $s_i$ are structurally unidentifiable. Fig \ref{fig1} illustrates this fact by showing that changes in the model outputs (i.e., glucose concentration and $\beta$-cell mass) resulting from halving the value of $s_i$ can be compensated by doubling the value of $p$. Therefore, it is not possible to distinguish between two parameter vectors of the form $(s_i,p)$ and $(s_i/2,2\cdot p)$.
This also entails that insulin is an unobservable state, since the impossibility of determining the true parameter vector leads to the impossibility of determining which of the time courses shown in the lower left plot of Fig \ref{fig1} is the true one. Therefore, the model cannot be used for inferring insulin concentration from measurements of the other variables. This limitation can be overcome if the value of $p$ or of $s_i$ is known.

Such lack of structural identifiability can have important consequences. A nice illustration is given in a recent work \cite{procopio2017model}, where Procopio et al presented a model of the release of a cardiac damage biomarker, cardiac troponin T, with the purpose of diagnosing acute myocardial infarction in a clinical setting. After the authors realized that the first version of the model was structurally unidentifiable, which could potentially lead to wrong conclusions, they removed the redundancies in their model and obtained an equivalent one that was structurally identifiable.

Structural unidentifiability is related to unobservability, as shown in the $\beta$IG model example, in which the inability to estimate $p$ leads to wrong predictions of $I$. However, unidentifiability does not always entail unobservability. As a trivial example, consider the case in which the value of $p$ is known. Then the $\beta$IG model becomes structurally identifiable and observable. If we now modify the model by replacing parameter $c$ with the sum of two new parameters ($c \rightarrow c_1 + c_2$), the two new parameters would obviously be structurally unidentifiable, but the unmeasured state $I$ would remain observable. Therefore, it is desirable to analyse both the structural identifiability and observability of a model to decrease the possibility of drawing false conclusions from it.

Before concluding this section, it should be noted that a structurally identifiable model may nevertheless be practically unidentifiable, that is, the numerical estimates of its parameters may contain large errors due to insufficient or bad quality data. A recent example of this scenario is given in \cite{eisenberg2017confidence}, where different models of cancer chemotherapy were analysed. The results showed that, although the models were structurally identifiable, they were not practically identifiable. This deficiency could lead to infer incorrect cell cycle distributions and, as a result, to the choice of suboptimal therapies.
It is thus reasonable to ask: if a model can be structurally identifiable and yet unidentifiable in practice, why should we care about analysing its structural identifiability in the first place? The answer is that practical and structural unidentifiability have different causes and also different remedies. Practical unidentifiability may be surmounted by using more informative data for calibration, but structural unidentifiabilities cannot be removed in this way (unless the new data involves modifying the output of the model, which strictly speaking entails modifying the model structure). Any attempt to remove a structural unidentifiability by incorporating more experimental data to the calibration (e.g. by sampling more densely or for a longer time) is doomed to fail, leading to a loss of resources and time. 
Practical identifiability analysis is not covered in this review; the interested reader is referred to \cite{distefano2015dynamic,walter1997identification,villaverde2016identifiability}.

In summary, it is advisable to analyse the observability and structural identifiability of a model before attempting to obtain insights from it.
If this analysis reveals deficiencies, actions must be taken depending on the intended application of the model.

For example, if the intended application is for determining the value of a parameter that turns out to be structurally unidentifiable, it is necessary to eliminate this structural identifiability. There are several ways of achieving this. Sometimes it may be possible to determine the unidentifiable parameter by direct measurements, either of the parameter of interest or of the parameter(s) that are correlated with it. However, direct measurements of parameters are seldom possible. It is often more practical to measure additional state variables, which may make the model (or at least the parameter of interest) structurally identifiable; this possibility should be analysed before performing the experiments. Finally, if the experimental setup cannot be modified, or if it is not practical to obtain new experimental data, one can try to modify the model structure by reducing the number of parameters. This can be achieved by fixing some parameters to values taken from the literature or by merging several unidentifiable parameters into an identifiable one.

If the intended application of the model is for determining the system states, as opposed to the parameters, a structurally unidentifiable model may still be useful -- as mentioned previously -- as long as the states of interest are observable. In this case, lack of observability may be remedied in a similar way as structural identifiability.

\section{Background: observability and structural identifiability}\label{sec:struct}

To define observability it is necessary to introduce the notion of distinguishable states:
\begin{definition}
Let $M$ be a model with internal state $x$ and measurable output $y$. Let $y_{x_0}(t)$ denote the time evolution of the model output when started from an initial state $x_0$ at $t_0$. Two states $x_1$ and $x_2$ are \textit{indistinguishable} if $y_{x_1}(t) = y_{x_2}(t)$ for all $t \geq t_0$. The set of states that are indistinguishable from $x_1$ is denoted by $I(x_1)$.
\end{definition}

A model is observable if it is possible to distinguish its internal state from any other state, that is:

\begin{definition}
	A model $M$ is \textit{observable} at $x_0$ if $I(x_0)=x_0$.
\end{definition}

Observability describes the possibility of determining the current state from present and future measurements. A similar concept, reconstructability, refers to determining the current state from present and past measurements.

\subsection{Observability of linear systems}

For illustration purposes, this subsection presents the special case of linear time invariant (LTI) systems, whose equations can be written as:

\begin{equation}\label{eq:L_model}
M_{L}:\left\{%
\begin{array}{lcl}
\dot{x}(t) & = & A(\theta)\cdot x(t)+B(\theta)\cdot u(t),\\
y(t)       & = & C(\theta)\cdot x(t),\\
x_0        & = & x(t_0,\theta)
\end{array}%
\right.
\end{equation}
where $\theta\in\mathbb{R}^q$ is the parameter vector, $u(t)\in\mathbb{R}^r$ the input vector, $x(t)\in\mathbb{R}^n$ the state variable vector, and $y(t)\in\mathbb{R}^m$ the output vector.
$A(\theta)$, $B(\theta)$, and $C(\theta)$ are constant matrices of dimensions $n\times n$, $n\times r$, and $m\times n$, respectively.
The dependence on $\theta$ may be dropped for ease of notation. 

Assessing the observability of $M_{L}$ amounts to determining whether it is possible to infer its internal state, $x$, by observing its output, $y$. An intuitive way of obtaining a condition for checking observability is the following. The available knowledge consists of the output and its derivatives, that is:

\begin{align}
\begin{array}{ll}
y         & = C\cdot x  \\
\dot y    & = C\cdot \dot x  =  C\cdot A\cdot x + C\cdot B\cdot u\\
\ddot y   & = C\cdot A\cdot \dot x + C\cdot B\cdot \dot u = \\
          & = C\cdot A^2\cdot x + C\cdot A\cdot B\cdot u + C\cdot B\cdot \dot u   \\
          & \vdots \\
\frac{d^i y}{dt^i} & = C\cdot A^i\cdot x + h\left(A,B,C,u,\dot u, \ddot u, \dots,\frac{ d^{i-1}u}{ dt^{i-1}}\right)\\                    
\end{array}%
\end{align}
where $h$ is a known matrix function.
Setting $i=n$ and writing the above equations in matrix form leads to
\begin{equation}\label{eq:int}
\begin{array}{l}
	\left(
	\begin{array}{c}
	y            \\
	\dot y     \\
	\ddot y   \\
	\vdots       \\
	\frac{d^{(n-1)} y}{dt^{(n-1)}}      
	\end{array}
	\right) = \left(
	\begin{array}{c}
	C            \\
	C\cdot A     \\
	C\cdot A^2   \\
	\vdots       \\
	C\cdot A^{n-1}      
	\end{array}
	\right) \cdot x + \ldots \\ \ldots + h\left(A,B,C,u,\dot u, \ddot u, \dots,\frac{ d^{n-2}u}{ dt^{n-2}}\right) = \\
	= {\mathcal O}^L \cdot x + h\left(A,B,C,u,\dot u, \ddot u, \dots,\frac{ d^{n-2}u}{ dt^{n-2}}\right)
\end{array}
\end{equation}
where the linear observability matrix has been introduced, ${\mathcal O}^L = \left(C|C\cdot A|C\cdot A^2|\dots|C\cdot A^{n-1}\right)^T$.
If ${\mathcal O}^L$ is invertible, one can uniquely obtain $x$ from the knowledge of $y$ and its derivatives, as long as $\text{rank}({\mathcal O}^L) = n$. This is known as the linear observability rank condition.
\begin{theorem}\label{LORC}{Linear Observability Rank Condition}.
	Given a linear time invariant model $M_L$ as defined in (\ref{eq:L_model}), a necessary and sufficient condition for complete observability is that $\text{rank}({\mathcal O}^L) = n$, where ${\mathcal O}^L = \left(C|C\cdot A|C\cdot A^2|\dots|C\cdot A^{n-1}\right)^T$ \cite{kalman1960general}.
\end{theorem}
``Complete'' observability means that all the model states can be inferred from observations of the output.

\subsection{Observability of nonlinear systems}\label{sec:NLobs}

Let us now consider nonlinear ODE models. In their most general form they can be written as:
\begin{equation}\label{eq:NL_model}
M_{NL}:\left\{%
\begin{array}{lcl}
\dot{x}(t) & = & f(x(t),u(t),\theta),\\
y(t) & = & g(x(t),\theta),\\
x_0 & = & x(t_0,\theta)
\end{array}%
\right.
\end{equation}
where $f$ and $g$ are analytic vector functions.
A special case of (\ref{eq:NL_model}) is that of nonlinear affine-in-the-input systems:
\begin{equation}\label{eq:affine_model}
M_{\text{aff}}:\left\{%
\begin{array}{lcl}
\dot{x}(t) & = & f_1(x(t),\theta)+f_2(x(t),\theta)\cdot u(t),\\
y(t)       & = & g(x(t),\theta),\\
x_0        & = & x(t_0,\theta)
\end{array}%
\right.
\end{equation}

Shortly after Kalman's introduction of the concept of observability \cite{kalman1960contributions,kalman1960general}, several researchers worked on its application to nonlinear systems of the type defined in equations (\ref{eq:NL_model}, \ref{eq:affine_model}). As a result, sufficient and/or necessary conditions for nonlinear observability were obtained \cite{kostyukovskii1968simple,griffith1971observability,kou1973observability,hermann1977nonlinear}, allowing to extend the observability rank condition in this context.
For nonlinear models, unlike for LTI models like (\ref{eq:L_model}), the derivatives of the output cannot be expressed in terms of the $A, B, C$ arrays. It is therefore necessary to define a nonlinear version of the observability matrix, ${\mathcal O}^{NL}$; to this end Lie derivatives are used.

\begin{definition}
	The \textit{Lie derivative} of $g(x)$ with respect to $f(x)$ is defined by:	 
	\begin{equation}\label{Lie1}
	L_f g(x) = \frac{\partial g(x)}{\partial x}f(x).
	\end{equation}	 
	Higher order Lie derivatives can be recursively calculated as:	 
	\begin{align}\label{Lie2}
	\begin{array}{rcl}
	L_f^2 g(x) & = & \frac{\partial L_f g(x)}{\partial x}f(x), \\
	& \vdots & \\
	L_f^i g(x) & = & \frac{\partial L_f^{i-1} g(x)}{\partial x}f(x).
	\end{array}
	\end{align}	
\end{definition}

It can be noticed from (\ref{eq:int}) that the linear observability matrix, ${\mathcal O}^{L}$, is the partial derivative of the derivatives of the output with respect to the states, that is,
\begin{align}\label{nonlinobs0}
{\mathcal O}^L = 
\left(
\begin{array}{c}
C            \\
C\cdot A     \\
C\cdot A^2   \\
\vdots       \\
C\cdot A^{n-1}      
\end{array}
\right) = 
\frac{\partial }{\partial x} \left(  
\begin{array}{c}
y        \\
\dot y   \\
\ddot y  \\
\vdots       \\
y^{(n-1)}\\      
\end{array}
\right) 
\end{align}

In a nonlinear model such as (\ref{eq:NL_model}) with constant input, $u(t)=u$, the $i^{th}$ Lie derivative of the output function $g(x)$ coincides with the $i^{th}$ time derivative of $y(t)$, i.e. $y^{(i)}(t)=L_f^i g(x)$. 
Thus, Lie derivatives can be used to calculate ${\mathcal O}^{NL}$ for nonlinear models with constant inputs as follows:

\begin{align}\label{nonlinobs}
{\mathcal O}^{NL}(x) = \left(  
\begin{array}{c}
\frac{\partial }{\partial x} y(t)        \\
\frac{\partial }{\partial x} \dot y(t)   \\
\frac{\partial }{\partial x} \ddot y(t)  \\
\vdots       \\
\frac{\partial }{\partial x} y^{(n-1)}(t)\\      
\end{array}
\right) 
= \left(  
\begin{array}{c}
\frac{\partial }{\partial x}g(x)           \\
\frac{\partial }{\partial x}(L_f g(x))     \\
\frac{\partial }{\partial x}(L_f^2 g(x))   \\
\vdots       \\
\frac{\partial }{\partial x}(L_f^{n-1}g(x))\\      
\end{array}
\right)
\end{align}

The nonlinear version of the observability rank condition can be stated as follows:

\begin{theorem}\label{NLORC}
	Nonlinear Observability Rank Condition: 
	If the model $M_{NL}$ given by (\ref{eq:NL_model}) with constant input $u$ satisfies $\text{rank}({\mathcal O}^{NL}(x_0)) = n$, where ${\mathcal O}^{NL}$ is defined by (\ref{nonlinobs}), then it is (locally) observable around $x_0$ \cite{hermann1977nonlinear,vidyasagar1993nonlinear}.
\end{theorem}
Two remarks are in order. First, it should be noted that the nonlinear observability rank condition (ORC) is a \textit{sufficient}, but not strictly necessary, condition for nonlinear observability (unlike the linear case, in which the ORC is both sufficient and necessary). In the nonlinear case, the ORC is ``almost necessary'' in the sense that, if $M_{NL}$ is locally observable around $x_0$, then $\text{rank}({\mathcal O}^{NL}(x_0)) = n$ for an open dense subset of the state space \cite{vidyasagar1993nonlinear}. This is a rather technical distinction, and in practice a failure to comply with the ORC is often considered as a very strong indication of unobservability.
Second, it should also be noted that the ORC determines \textit{local} observability: if a model satisfies the ORC, it is possible to distinguish between two adjacent states, but there may still be distant states that are indistinguishable. A locally observable model is often -- although not always -- globally observable too.

\subsection{Structural local identifiability as observability}\label{sec:SI}

In this paper structural identifiability is considered as a particular case of observability. As noted in the preceding subsection \ref{sec:NLobs}, nonlinear observability is a local concept, which means we will study structural \textit{local} identifiability. The analysis of structural \textit{global} identifiability requires other approaches \cite{Chis11a,miao2011identifiability,villaverde2016identifiability}. Note however that the definitions provided here do not prevent a locally identifiable model to be also globally identifiable, and this will actually be the case in many practical applications.

\begin{definition}
	A parameter $\theta_i$ in a model $M_{NL}$ given by (\ref{eq:NL_model}) is \textit{structurally locally identifiable} (s.l.i.) if for almost any parameter vector $\theta^*\in\mathbb{R}^q$ there is a neighbourhood ${\mathcal N}(\theta^*)$ such that the following property holds:
\begin{equation}\label{eq:sli}
\hat{\theta} \in {\mathcal N}(\theta^*) \text{ and } g(x,\hat{\theta}) = g(x,\theta^*) \Rightarrow \hat{\theta_i} = \theta_i^*
\end{equation}
\end{definition}

\begin{definition}
A parameter $\theta_i$ is \textit{structurally unidentifiable} (s.u.) if (\ref{eq:sli}) does not hold in any neighbourhood of $\theta^*$.
\end{definition}

\begin{definition}
	A model $M_{NL}$ is s.l.i. if all its parameters are s.l.i.. 
\end{definition}

\begin{definition}
A model $M_{NL}$ is s.u. if at least one of its parameters is s.u..
\end{definition}

Structural identifiability can be considered as a particular case of observability by considering the parameters as state variables with zero dynamics \cite{tunali1987new,sedoglavic2002probabilistic,anguelova2004nonlinear,anguelova2007observability,august2009new}. The augmented state variable vector is:
\begin{align}
\tilde{x} = \left[
\begin{array}{c}
x \\
\theta \\
\end{array}
\right]
\end{align}
Similarly to the nonlinear observability matrix of (\ref{nonlinobs}), it is possible to define an augmented nonlinear observability-identifiability matrix, ${\mathcal O}^{NL}_I(\tilde{x})$, as:

\begin{align}\label{obsident}
{\mathcal O}^{NL}_I(\tilde{x} ) = \left(
\begin{array}{c}
\frac{\partial }{\partial \tilde{x}}g(\tilde{x})             \\
\frac{\partial }{\partial \tilde{x}}(L_f g(\tilde{x}))       \\
\frac{\partial }{\partial \tilde{x}}(L_f^2 g(\tilde{x}))     \\
\vdots       \\
\frac{\partial }{\partial \tilde{x}}(L_f^{n+q-1}g(\tilde{x}))\\      
\end{array}
\right)
\end{align}

\begin{theorem}\label{OIC} Nonlinear Observability-Identifiability Condition (OIC).
	If a model $M_{NL}$ given by (\ref{eq:NL_model}) satisfies $\text{rank}({\mathcal O}^{NL}_I(\tilde x_0)) = n+q$, with ${\mathcal O}^{NL}_I(\tilde x_0)$ given by (\ref{obsident}), then it is (locally) observable and identifiable in a neighbourhood ${\mathcal N}(\tilde x_0)$ of $\tilde x_0$.  
\end{theorem}

\begin{remark}\label{remark1}	
	Identifiability of individual parameters: if the OIC condition is fulfilled, all the parameters of $M_{NL}$ are s.l.i.. If the OIC does not hold, $M_{NL}$ is s.u. and at least some parameter(s) are s.u. (and/or some states are unobservable). Since each column in ${\mathcal O}^{NL}_I$ corresponds to the partial derivative with respect to a state or parameter, it is possible to determine which parameters (states) are structurally unidentifiable (unobservable) by removing the corresponding column and recalculating rank(${\mathcal O}^{NL}_I$). If deleting the $i^{th}$ column does not change rank(${\mathcal O}^{NL}_I$), then the $i^{th}$ parameter (state) is structurally unidentifiable (unobservable) \cite{anguelova2004nonlinear}.
	We can thus define a Structural Identifiability Condition for a parameter as follows:
\end{remark}

\begin{theorem}\label{SIC} Structural Identifiability Condition (SIC).
	Given a model $M_{NL}$ defined by (\ref{eq:NL_model}), its $i^{th}$ parameter $\theta_i$ is structurally locally identifiable in a neighbourhood ${\mathcal N}(\tilde x_0)$ of $\tilde x_0$ if $\text{rank}({\mathcal O}^{i*}_I(\tilde x_0)) < \text{rank}({\mathcal O}_I(\tilde x_0))$, where ${\mathcal O}_I(\tilde x_0)$ is the ${\mathcal O}^{NL}_I(\tilde x_0)$ defined in (\ref{obsident}), and ${\mathcal O}^{i*}_I(\tilde x_0)$ is the array that results from removing the column corresponding to $\partial/\partial \theta_i$ from ${\mathcal O}_I(\tilde x_0)$.  
\end{theorem}

\subsection{Example: observability and structural identifiability analysis of a nonlinear model}\label{sec:example}

The approach described in subsection \ref{sec:SI} is demonstrated here by applying it to the nonlinear model used as motivating example in section \ref{sec:motivation}. This case study was briefly described in Section \ref{sec:motivation} and Fig \ref{fig1}, which shows its dynamic equations. It consists of $n=3$ states, $x = [G,\beta,I]$, $m=2$ outputs, $y = [G, \beta]$, $q=5$ parameters, $\theta = [p,s_i,\gamma,c,\alpha]$, and $r=1$ input, $u$. The augmented vector consisting of the states and parameters is this $\tilde x = [G,\beta,I,p,s_i,\gamma,c,\alpha]$.

The observability and structural identifiability of this system can be analysed with the observability-identifiability condition (OIC) of Theorem \ref{OIC}. To this end one must build the ${\mathcal O}^{NL}_I$ matrix defined in equation (\ref{obsident}).
The first two rows in ${\mathcal O}^{NL}_I$ correspond to the partial derivatives of the output function with respect to the states and parameters; since the output is $y = g(\tilde x) = [G, \beta]$, the first two rows of ${\mathcal O}^{NL}_I$ are:

\begin{align*}
\frac{\partial G}{\partial \tilde x} = [1,0,0,0,0,0,0]\\
\frac{\partial \beta}{\partial \tilde x} = [0,1,0,0,0,0,0] 
\end{align*}

The matrix made up of the two rows above has rank equal to two.
Subsequent rows are calculated with Lie derivatives as defined in equations (\ref{Lie1}, \ref{Lie2}). In principle, $n+q-1=7$ Lie derivatives must be symbolically calculated. However, in practice it may be possible to stop the calculation earlier: if the rank of the matrix does not increase after the addition of a new derivative, it is not necessary to calculate higher order derivatives since they will not modify the rank. 

The first Lie derivative is obtained as:
\begin{flalign*}
L_f g(\tilde{x}) = \frac{\partial g(\tilde x)}{\partial \tilde x}f(\tilde x) = \\
\left(  
\begin{array}{c}
u + u_0 - x_1\cdot(p_4 + p_2\cdot x_3) \\ 
x_2\cdot\left( \frac{1.4583\cdot 10^{-5}}{ \left(\frac{8.4}{x_1}\right)^{1.7} + 1} - \frac{1.7361\cdot 10^{-5}}{\left(\frac{x_1}{4.8}\right)^{8.5} + 1}   \right)  
\end{array}
\right) 
\end{flalign*}

Thus, the third and fourth rows of ${\mathcal O}^{NL}_I$ are:

\begin{flalign*}
\frac{\partial }{\partial \tilde{x}}(L_f g(\tilde{x})) =  \\
\left(  
\begin{array}{cccccccc}
{\cal O}_{3,1} & 0 & {\cal O}_{3,3} & 0 & {\cal O}_{3,5} & 0 & {\cal O}_{3,7} & 0 \\
{\cal O}_{4,1}  & {\cal O}_{4,2}  & 0 & 0 & 0 & 0 & 0 & 0
\end{array} 
\right)  
\end{flalign*}

\noindent where 

\begin{flalign*}
\begin{array}{c}
{\cal O}_{3,1} = -c-s_i\cdot x_3\\
{\cal O}_{3,3} = -s_i\cdot x_1\\
{\cal O}_{3,5} = -x_1\cdot x_3\\
{\cal O}_{3,7} = -x_1\\
{\cal O}_{4,1} = \frac{3.0743\cdot 10^{-5}\cdot x_2\cdot\left(\frac{x_1}{4.8}\right)^{7.5}} 
{\left(   \left(\frac{x_1}{4.8}\right)^{8.5} + 1 \right)^2 }    
- \frac{0.0052\cdot \left(\frac{8.4}{x_1}\right) ^{0.7} }
{\left( 25\cdot x_1^2 \cdot\left(\frac{8.4}{x_1}\right)^{1.7} + 1  \right)^2 } \\
{\cal O}_{4,2} = \frac{1.4583\cdot 10^{-5}}{ \left(\frac{8.4}{x_1}\right)^{1.7} + 1} - \frac{1.7361\cdot 10^{-5}}{\left(\frac{x_1}{4.8}\right)^{8.5} + 1}    
\end{array} 
\end{flalign*}

By adding the two rows corresponding to $\frac{\partial }{\partial \tilde{x}}(L_f g(\tilde{x}))$, the rank of ${\mathcal O}^{NL}_I$ increases from two to three. Proceeding in the same manner, the rank of the matrix increases with every additional Lie derivative until it stops: it is equal to 7 when ${\mathcal O}^{NL}_I$ is built with both 5 and 6 Lie derivatives. Thus with 6 derivatives we know that the model has some observability/identifiability issues, since its matrix does not have full rank.

At this point we can determine the observability of each state and the structural identifiability of each parameter using the procedure described in Remark \ref{remark1}. This yields that the unmeasured state $I$ is not observable, and that there are two s.u. parameters ($p,s_i$) and three s.l.i. parameters ($\gamma,c,\alpha$).
It can be noticed that multiplying by $s_i$ the dynamic equation of $I$ shown in Fig 1 leads to a modified model in which the third state is $(s_i\cdot I)$ instead of $I$, and parameter $p$ only appears in the equations as part of the product $s_i\cdot p$. This model formulation highlights the fact that only the products $s_i\cdot p$ and $s_i\cdot I$ are observable (identifiable).

\section{Recent developments}\label{sec:recent}

\subsection{Computational implementations of the rank conditions}

The conditions described in Section \ref{sec:struct} involve building observability (${\mathcal O}^{NL}$) or observability-identifiability matrices (${\mathcal O}^{NL}_I$) and calculating their rank. Building these arrays involves symbolic calculations, which can be performed in environments such as Mathematica (Wolfram Research, Champaign, IL, USA), MATLAB (MathWorks, Natick, MA, USA), or MAPLE (Maplesoft, Waterloo, ON, Canada). Some software tools provide advanced implementations of these calculations.

August and Papachristodoulou \cite{august2009new} used semidefinite programming to evaluate the OIC (Theorem \ref{OIC}). They used SOSTOOLS \cite{prajna2002introducing}, a free MATLAB toolbox that performs a sum of squares decomposition. This technique allows to assess identifiability for all parameter values within an interval; however, the computational cost of the rank calculation quickly becomes high as the problem size increases, which hinders the applicability of this method to medium-to-large models. 

Another MATLAB tool is the STRIKE-GOLDD toolbox \cite{villaverde2016structural}, a publicly available software that analyses structural identifiability and observability using the OIC. It includes options such as performing partial analyses and decomposing the models, which can be helpful for analysing large models.

For rational systems, the Exact Arithmetic Rank (EAR) method is a numerical alternative for calculating the rank. It is based on an algorithm originally presented by Sedoglavic \cite{sedoglavic2002probabilistic}, which was extended and implemented in Mathematica by Jirstrand and coworkers \cite{anguelova2012efficient}.

\subsection{Accessibility and the role of initial conditions}

The rank conditions of Theorems \ref{NLORC}--\ref{OIC} provide results that are valid for ``almost all'' values of the variables (state and parameter vectors), that is, for all possible values except for a set of measure zero (a ``thin set'').
Consequently, for specific values there may be loss of identifiability. This was pointed out by Saccomani et al. \cite{saccomani2003parameter,Dangio2009identifiability}, who analysed this phenomenon with a differential algebra approach, tracing its cause to a loss of accessibility from certain initial conditions.
Accessibility, also called reachability, is a property that describes the ability to move a system to any state in a neighbourhood of the initial one.
Saccomani and coworkers noted that a loss of accessibility from specific initial conditions could lead to loss of structural identifiability. 

This matter has been recently approached from the differential geometry viewpoint. In \cite{villaverde2017structural} it was remarked that loss of accessibility is not the only possible cause of loss of structural identifiability from specific initial conditions: this phenomenon can take place even for models that are not accessible from generic initial conditions. Furthermore, it was also noted that a decrease in rank(${\mathcal O}^{NL}_I$) at a specific initial condition $x(0)$ does not necessarily result in a loss of structural identifiability, even if the system is started at that initial condition.
In \cite{villaverde2017structural} a method for finding potentially problematic vectors was also suggested, although it scales up poorly with system size.

\subsection{The role of inputs}

The methodology presented in Section \ref{sec:struct} assumes that the input vector $u$ is \textit{known and constant}. Obviously, the same formulation can account for the case of \textit{unknown constant} inputs simply by considering them as additional parameters -- which are unknown and constant by definition. 
For \textit{known}, \textit{time-varying} inputs that are differentiable functions of time, a differential algebra approach would still be valid. However, the differential geometry procedure described in Section \ref{sec:struct} needs to be extended in order to cope with this case. To this end it has recently been suggested to use Extended Lie derivatives \cite{villaverde2018input}, which are defined as follows:

\begin{definition}
	The \textit{extended Lie derivative} is \cite{anguelova2012efficient}:
	\begin{equation}\label{Lie_ext1}
	L_f g(\tilde{x}) = \frac{\partial g(\tilde{x})}{\partial \tilde{x}}f(\tilde{x},u) + \sum_{j=0}^{j=\infty}\frac{\partial g(\tilde{x})}{\partial u^{(j)}}u^{(j+1)}
	\end{equation}	
	where $u^{(j)}$ is the $j^{th}$ derivative of the input $u$.
	Higher order extended Lie derivatives are recursively calculated as:
	\begin{equation}\label{Lie_ext2}
	L_f^i g(\tilde{x}) = \frac{\partial L_f^{i-1} g(\tilde{x})}{\partial \tilde{x}}f(\tilde{x},u) + \sum_{j=0}^{j=\infty}\frac{\partial L_f^{i-1} g(\tilde{x})}{\partial u^{(j)}}u^{(j+1)}
	\end{equation}
\end{definition}
(Note that this definition considers a time-dependent input vector $u(t)$, which is simply written as $u$ for ease of notation.)
Unlike the previously defined Lie derivatives of (\ref{Lie1}, \ref{Lie2}), the extended Lie derivatives are equal to the output derivatives for time-varying inputs, $y^{(i)}(t)=L_f^ig(x)$.
Evaluating the OIC with a $\mathcal O^{NL}_I$ built with extended Lie derivatives correctly determines the observability and structural identifiability of a model. 
Some models may require time-varying inputs in order to be identifiable. In \cite{villaverde2018input} it was shown how the extended Lie derivatives can be used for experimental design, by determining the number of non-zero derivatives of the input that are required for structural identifiability. 

The identifiability of the $\beta$IG model used in sections \ref{sec:motivation} and \ref{sec:example} does not depend on the input derivatives. Hence in this section this situation will be illustrated with a different example, the following two-compartment model \cite{villaverde2018input}:

\begin{equation}\label{eq:comp}
\begin{array}{ll}
\dot{x_1} &= -(k_{1e} + k_{12})\cdot x_1 + k_{21}\cdot x_2 + b\cdot u,\\
\dot{x_2} &= k_{12}\cdot x_1 - k_{21}\cdot x_2,\\
y         &=  x_1
\end{array}
\end{equation}
Compartmental models of this type are commonly used to describe physiological processes. 
Note that, although the model given by (\ref{eq:comp}) is linear in the states, if the state vector is augmented with the parameters (as needed for structural identifiability analysis) the model becomes nonlinear.

\begin{figure}[h]
	\begin{center}
		\includegraphics[width=1\linewidth]{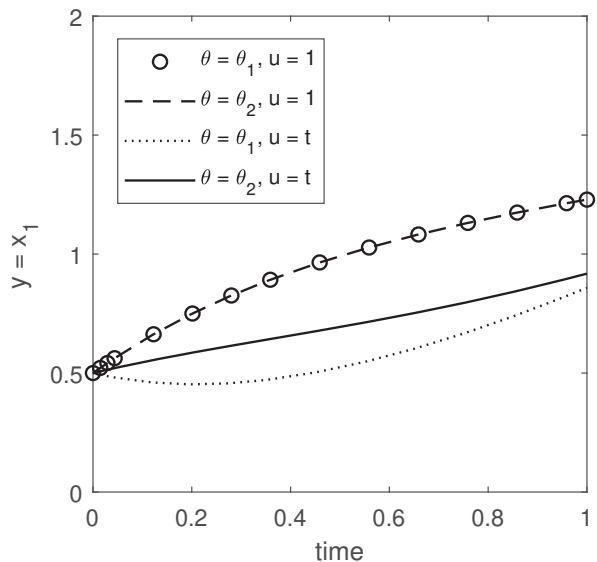}
		\caption{Output of the two-compartment model of equations (\ref{eq:comp}) for two different parameter vectors ($\theta_1$ given by $k_{1e} = 1, k_{12} = 3, k_{21} = 1, b = 1$, and $\theta_2$ given by $k_{1e} = 2, k_{12} = 2.5, k_{21} = 0.5, b = 2$), and two different inputs, $u=1$ and $u=t$ (where $t$ stands for time). With a constant input $u=1$ both parameter vectors are indistinguishable from the model output (there is actually an infinite number of pairs of indistinguishable parameter vectors), and the parameters are structurally unidentifiable. However, with a ramp input $u=t$ two different parameter vectors yield two different model outputs; in this case the parameters are structurally identifiable.}
		\label{fig:two-compart} 
	\end{center}
\end{figure}

This model is structurally unidentifiable from an experiment with a constant input, but becomes structurally identifiable with a continuous time-varying input such as a ramp \cite{villaverde2018input}. 
This is illustrated in Fig \ref{fig:two-compart}.
The constant input result can be obtained by applying the procedure described in Section \ref{sec:SI} as shown in Section \ref{sec:example}. Since this model has $n=2$ states and $q=4$ parameters, it would require $\text{rank}({\mathcal O}^{NL}_I)=n+q=6$ to be observable and s.l.i. However, the aforementioned procedure yields $\text{rank}({\mathcal O}^{NL}_I)=5$, and the procedure in Remark \ref{remark1} determines that $x_2$ is observable but all the parameters are s.u.
The time-varying input result is obtained by building ${\mathcal O}^{NL}_I$ with the extended Lie derivatives defined in (\ref{Lie1}, \ref{Lie2}); in the corresponding symbolic derivations $\dot u$ is set to a constant value and higher order derivatives, $\ddot u, \dddot u, \dots$ are set to zero. This yields $\text{rank}({\mathcal O}^{NL}_I)=6$ with 5 derivatives, and the model is observable and s.l.i.
These calculations can be performed with STRIKE-GOLDD2 \cite{villaverde2018input} and take less than one second in a standard computer. 
The difference in the results with $\dot u=0$ and $\dot u\neq0$ is due to the presence of terms containing $\dot u$ in some entries of ${\mathcal O}^{NL}_I$, whose contribution is needed for a full rank. Setting $\dot u=0$ removes these terms and decreases the matrix rank, leading to a loss of identifiability.

It should be noted that this model can also be analysed with a differential algebra approach; for example, the COMBOS application \cite{meshkat2014finding} obtains the same result in comparable time. Compared to the differential geometry approach, the advantages of this method are the ability to distinguish between local and global identifiability and to find identifiable combinations. Its disadvantages are that in principle it cannot consider specific derivatives being zero (e.g. $\dot u\neq0$ but  $\ddot u=0$) and that it typically has worse computational scale-up for models with large nonlinearities.

A different problem arises when the inputs are \textit{time-varying} and \textit{unknown}. Such inputs can be viewed as external disturbances, of which there are no measurements nor information about their dependence on time.
Martinelli \cite{martinelli2015extension} extended the ORC to account for this situation for the case of nonlinear systems that are affine with respect to the inputs, which must be differentiable but may be known and/or unknown.
To this end, the model defined by (\ref{eq:affine_model}) is augmented in order to include an unknown input vector $w$ as follows:
\begin{equation}\label{eq:affine_model_unknown_input}
\left\{%
\begin{array}{lcl}
\dot{x} & = & f_1(x,\theta)+f_2(x,\theta)\cdot u + f_3(x,\theta)\cdot w,\\
y       & = & g(x,\theta),\\
\end{array}%
\right.
\end{equation}
In \cite{martinelli2015extension} it was proposed to extend this model by augmenting the original state $x$ to $^kx$, which includes the input and its derivatives up to order $k$, that is $^kx = [x,\theta,\dot w, \ddot w, \dots, w^{(k)}]$. An extended observability rank condition (EORC) was then presented, allowing to check the observability of systems with unknown inputs, although not of the inputs themselves, at least in its published form. Although in \cite{martinelli2015extension,martinelli2018nonlinear} the structural identifiability problem was not explicitly considered, it is of course possible to apply this idea to a joint observability and structural identifiability analysis.

\subsection{Model symmetries and identifiable combinations}

If a set of parameters are found to be structurally unidentifiable, a question naturally arises: it is possible to reformulate the model by combining such parameters in an identifiable quantity?
The answer to this question entails characterizing the form in which the structurally unidentifiable parameters are correlated. Many methods for structural identifiability analysis are capable of addressing this problem to a certain extent; however, no generally applicable and automatic procedure exists.  

One of the first examples, the ``exhaustive modeling'' method for finding the set of models that are output indistinguishable from a given one, was presented in  \cite{walter1981unidentifiable}. This procedure, also known as the similarity transformation approach, can be used to obtain structurally identifiable versions of linear compartmental models. An extension to controlled nonlinear models, which requires testing controllability and observability conditions, was presented in \cite{vajda1989similarity}, and the case of uncontrolled systems was considered in \cite{joly1998technical,evans2002identifiability}.

Differential algebra is a classic approach for the study of observability \cite{diop1991nonlinear} and structural identifiability \cite{ljung1994global}. The equivalence between the observability definitions from the algebraic and differential geometric viewpoints was established in \cite{diop1993equivalence} for a class of rational systems.
DAISY is a software that adopts the differential algebra approach to assess global structural identifiability and observability \cite{bellu2007daisy}, and COMBOS \cite{meshkat2014finding} is a tool specifically developed for finding identifiable parameter combinations using differential algebra concepts such as Gr{\"o}bner bases \cite{meshkat2009algorithm,meshkat2011finding}.

Other approaches to this problem use Lie transformations. A method based on the generation of Lie algebras that represent the symmetries of the model equations was presented in \cite{yates2009structural}. This procedure uses random numerical specializations and is valid for autonomous, rational systems. 
Instead of using random specializations, another method described in \cite{merkt2015higher} finds Lie symmetries by transforming rational terms into linear terms.
Finally, the aforementioned toolbox STRIKE-GOLDD \cite{villaverde2016structural}, which uses Lie derivatives to calculate the observability-identifiability matrix ${\mathcal O^{NL}_I}$, includes a procedure for finding identifiable parameter combinations that is based on ideas from \cite{chappell1998procedure,evans2000extensions,anguelova2004nonlinear}. Briefly, it removes from ${\mathcal O^{NL}_I}$ the columns corresponding to identifiable parameters and calculates a basis for the null space of the resulting matrix. The coefficients of this basis define a set of partial differential equations, whose solutions yield the identifiable combinations.

\subsection{Sloppiness, dynamical compensation, and structural identifiability}

A structurally unidentifiable model can yield the same output for different parameter values. This situation might be interpreted as a sign of robustness of the system to changes in parameter values. However, while lack of identifiability is usually considered an undesirable model property, in certain contexts robustness is seen as a desirable property. This apparent contradiction highlights the subtle character of the relationship between identifiability and robustness. As an illustration of this relationship, this subsection discusses two concepts developed in recent years -- sloppiness and dynamical compensation -- that are related but not equivalent to unidentifiability.

The first concept, sloppiness or sloppy models, was introduced in \cite{brown2003statistical} to refer to the situation in which the model output is sensitive to changes in so-called stiff parameters, but largely insensitive to changes in sloppy parameters. Sloppiness was defined as the existence of a clear gap between the eigenvalues of the system's Fisher information matrix (FIM), with large eigenvalues corresponding to stiff parameters and small eigenvalues corresponding to sloppy parameters. It was claimed that sloppiness is a universal feature of systems biology models \cite{gutenkunst2007universally}, which would make it impossible to estimate all parameters accurately. 
More recent publications have provided new insights about sloppiness, as reviewed in \cite{transtrum2015perspective}.
The concept of sloppiness, which has been linked to information theory, highlights the fact that a model's output behaviour may still be tightly constrained despite the parameter values being only loosely constrained. Sloppiness provides a viewpoint for studying how distinguishable models are, and how they can be reduced.
Several papers have clarified the relation between sloppiness and identifiability \cite{apgar2010sloppy,tonsing2014cause,raman2017delineating,chis2016relationship}. 
It is now understood that sloppiness is related to \textit{practical} rather than \textit{structural} identifiability, and that it is not equivalent to unidentifiability of any kind, meaning that sloppy models can indeed be identifiable. 

The second concept, dynamical compensation (DC for short), was introduced in \cite{karin} as a property found in certain physiological circuits.
Originally DC was defined simply as the invariance of the model output with respect to changes in a parameter value. It was immediately noted that according to this definition DC amounted to structural unidentifiability \cite{sontag2017dynamic,villaverde2017dynamicalarxiv}. (Note that the glucose homeostasis mechanism discussed in the Introduction was proposed in \cite{karin} as a possible mechanism for achieving DC; depending on its formulation -- i.e. on which states are measured and which parameters are known -- this model can be structurally unidentifiable). This equivalence between structural unidentifiability and the original definition of DC was not discussed in \cite{karin} and was potentially problematic, since 
the purpose of DC was to describe a phenomenon different to structural unidentifiability. More precisely, DC referred to the capability of a physiological circuit to maintain its dynamic behaviour unchanged after a change in the value of a model parameter, following a transition period. An alternative definition of DC that provided a more detailed description of the phenomenon and that took into account the relationship with structural identifiability was proposed in \cite{villaverde2017dynamical}.

\section{Open problems and future directions}\label{sec:conclusions}

The differential geometry approach adopted in this review has been used to analyse observability and structural identifiability of nonlinear systems for more than forty years. The theoretical and computational advances made in the last decades have increased its applicability. However, there are still many challenges that call for more research in this area. 

For example, an intrinsic limitation of the approach is that it yields only \textit{local} results. Other methods, such as differential algebra, are capable of providing global structural identifiability results. They could possibly serve as an inspiration for extending (hybridizing?) the differential geometry techniques to perform global analyses.

Other desirable developments would consist of advanced implementations to alleviate the computational burden of the analyses. Such improvements, which may benefit from the use of parallelization and high performance computing techniques, would facilitate the application of these methods to the increasingly large models being built in the biological modelling community. 

Another possible direction concerns the role of inputs in observability and identifiability analysis. Despite recent advances, there are still several open questions regarding this matter.
It has been noted that certain models that are structurally unidentifiable from a single constant input experiment can become identifiable if a continuously time-varying input is used \cite{villaverde2018input}. In some cases the same improvement can be obtained with multiple constant input experiments \cite{ligon2018genssi,villaverde2018input} -- or, equivalently, with a single experiment with a piecewise constant input. 
However, the question of when a time-varying input and multiple constant inputs are equivalent for the purpose of structural identifiability has not been answered yet.
Likewise, the problem of analysing observability and structural identifiability in presence of unmeasured inputs has not been fully solved yet.

Finally, an important open question is the relationship between observability/identifiability and model predictions. On the one hand, it is known that lack of the former can lead to errors in the latter. On the other hand, it is true that this is not necessarily the case. Therefore, further insights into the requisites for accurate predictive modelling would be a valuable contribution.

\section*{Acknowledgements}
The author was supported by the European Union's Horizon 2020 research and innovation programme under grant agreement No 686282 (``CANPATHPRO'') during the writing of this paper.


\begin{thebibliography}{10}
	
	\bibitem{anguelova2004nonlinear}
	M.~Anguelova.
	\newblock Nonlinear observability and identifiability: General theory and a
	case study of a kinetic model for \textit{S. cerevisiae}.
	\newblock Master's thesis, Chalmers University of Technology and G\"oteborg
	University, 2004.
	
	\bibitem{anguelova2007observability}
	M.~Anguelova.
	\newblock {\em Observability and identifiability of nonlinear systems with
		applications in biology}.
	\newblock PhD thesis, Chalmers University of Technology, 2007.
	
	\bibitem{apgar2010sloppy}
	J.~F. Apgar, D.~K. Witmer, F.~M. White, and B.~Tidor.
	\newblock Sloppy models, parameter uncertainty, and the role of experimental
	design.
	\newblock {\em Mol. Biosyst.}, 6(10):1890--1900, 2010.
	
	\bibitem{audoly2001global}
	S.~Audoly, G.~Bellu, L.~D'Angi{\`o}, M.~P. Saccomani, and C.~Cobelli.
	\newblock Global identifiability of nonlinear models of biological systems.
	\newblock {\em IEEE Trans. Biomed. Eng.}, 48(1):55--65, 2001.
	
	\bibitem{august2009new}
	E.~August and A.~Papachristodoulou.
	\newblock A new computational tool for establishing model parameter
	identifiability.
	\newblock {\em J. Comput. Biol.}, 16(6):875--885, 2009.
	
	\bibitem{bellman1970structural}
	R.~Bellman and K.~J. {\AA}str{\"o}m.
	\newblock On structural identifiability.
	\newblock {\em Math. Biosci.}, 7(3):329--339, 1970.
	
	\bibitem{bellu2007daisy}
	G.~Bellu, M.~P. Saccomani, S.~Audoly, and L.~D'Angi{\`o}.
	\newblock {DAISY}: a new software tool to test global identifiability of
	biological and physiological systems.
	\newblock {\em Comput. Methods Programs Biomed.}, 88(1):52--61, 2007.
	
	\bibitem{brown2003statistical}
	K.~S. Brown and J.~P. Sethna.
	\newblock Statistical mechanical approaches to models with many poorly known
	parameters.
	\newblock {\em Phys. Rev. E}, 68(2):021904, 2003.
	
	\bibitem{chappell1998procedure}
	M.~J. Chappell and R.~N. Gunn.
	\newblock A procedure for generating locally identifiable reparameterisations
	of unidentifiable non-linear systems by the similarity transformation
	approach.
	\newblock {\em Math. Biosci.}, 148(1):21--41, 1998.
	
	\bibitem{chatzis2015observability}
	M.~N. Chatzis, E.~N. Chatzi, and A.~W. Smyth.
	\newblock On the observability and identifiability of nonlinear structural and
	mechanical systems.
	\newblock {\em Struct. Control Hlth}, 22(3):574--593, 2015.
	
	\bibitem{Chis11b}
	O.-T. Chis, J.~R. Banga, and E.~Balsa-Canto.
	\newblock {GenSSI}: a software toolbox for structural identifiability analysis
	of biological models.
	\newblock {\em Bioinformatics}, 27(18):2610--2611, 2011.
	
	\bibitem{Chis11a}
	O.-T. Chis, J.~R. Banga, and E.~Balsa-Canto.
	\newblock Structural identifiability of systems biology models: a critical
	comparison of methods.
	\newblock {\em PLoS One}, 6(11):e27755, 2011.
	
	\bibitem{chis2016relationship}
	O.-T. Chis, A.~F. Villaverde, J.~R. Banga, and E.~Balsa-Canto.
	\newblock On the relationship between sloppiness and identifiability.
	\newblock {\em Math. Biosci.}, 282:147--161, 2016.
	
	\bibitem{Dangio2009identifiability}
	L.~D'Angi{\`o}, M.~P. Saccomani, S.~Audoly, and G.~Bellu.
	\newblock Identifiability of nonaccessible nonlinear systems.
	\newblock In R.~Bru and Romero-Viv{\'o}, editors, {\em Positive Systems},
	volume 389 of {\em LNCIS}, pages 269--277. Springer-Verlag, 2009.
	
	\bibitem{denis2001some}
	L.~Denis-Vidal, G.~Joly-Blanchard, and C.~Noiret.
	\newblock Some effective approaches to check the identifiability of
	uncontrolled nonlinear systems.
	\newblock {\em Math. Comput. Simul}, 57(1):35--44, 2001.
	
	\bibitem{diop1991nonlinearCDC}
	S.~Diop and M.~Fliess.
	\newblock Nonlinear observability, identifiability, and persistent
	trajectories.
	\newblock In {\em Proceedings of the 30th IEEE Conference on Decision and
		Control}, pages 714--719. IEEE, 1991.
	
	\bibitem{diop1991nonlinear}
	S.~Diop, M.~Fliess, et~al.
	\newblock On nonlinear observability.
	\newblock In {\em Proc. 1st Europ. Control Conf}, pages 152--157, 1991.
	
	\bibitem{diop1993equivalence}
	S.~Diop and Y.~Wang.
	\newblock Equivalence between algebraic observability and local generic
	observability.
	\newblock In {\em Proceedings of the 32nd IEEE Conference on Decision and
		Control}, pages 2864--2865. IEEE,
	1993.
	
	\bibitem{distefano2015dynamic}
	J.~DiStefano~III.
	\newblock {\em Dynamic systems biology modeling and simulation}.
	\newblock Academic Press, 2015.
	
	\bibitem{eisenberg2017confidence}
	M.~C. Eisenberg and H.~V. Jain.
	\newblock A confidence building exercise in data and identifiability: Modeling
	cancer chemotherapy as a case study.
	\newblock {\em J. Theor. Biol.}, 431:63--78, 2017.
	
	\bibitem{evans2002identifiability}
	N.~D. Evans, M.~J. Chapman, M.~J. Chappell, and K.~R. Godfrey.
	\newblock Identifiability of uncontrolled nonlinear rational systems.
	\newblock {\em Automatica}, 38(10):1799--1805, 2002.
	
	\bibitem{evans2000extensions}
	N.~D. Evans and M.~J. Chappell.
	\newblock Extensions to a procedure for generating locally identifiable
	reparameterisations of unidentifiable systems.
	\newblock {\em Math. Biosci.}, 168(2):137--159, 2000.
	
	\bibitem{griffith1971observability}
	E.~W. Griffith and K.~Kumar.
	\newblock On the observability of nonlinear systems: I.
	\newblock {\em J. Math. Anal. Appl.}, 35(1):135--147, 1971.
	
	\bibitem{gutenkunst2007universally}
	R.~N. Gutenkunst, J.~J. Waterfall, F.~P. Casey, K.~S. Brown, C.~R. Myers, and
	J.~P. Sethna.
	\newblock Universally sloppy parameter sensitivities in systems biology models.
	\newblock {\em PLoS Comput. Biol.}, 3(10):e189, 2007.
	
	\bibitem{hermann1977nonlinear}
	R.~Hermann and A.~J. Krener.
	\newblock Nonlinear controllability and observability.
	\newblock {\em IEEE Trans. Autom. Control}, 22(5):728--740, 1977.
	
	\bibitem{hong2018global}
	H.~Hong, A.~Ovchinnikov, G.~Pogudin, and C.~Yap.
	\newblock Global identifiability of differential models.
	\newblock {\em arXiv preprint arXiv:1801.08112}, 2018.
	
	\bibitem{isidori1995nonlinear}
	A.~Isidori.
	\newblock {\em Nonlinear control systems}.
	\newblock Springer Science \& Business Media, 1995.
	
	\bibitem{janzen2016parameter}
	D.~L. Janz{\'e}n, L.~Bergenholm, M.~Jirstrand, J.~Parkinson, J.~Yates, N.~D.
	Evans, and M.~J. Chappell.
	\newblock Parameter identifiability of fundamental pharmacodynamic models.
	\newblock {\em Front. Physiol}, 7, 2016.
	
	\bibitem{joly1998technical}
	G.~Joly-Blanchard and L.~Denis-Vidal.
	\newblock Technical communique: Some remarks about an identifiability result of
	nonlinear systems.
	\newblock {\em Automatica}, 34(9):1151--1152, 1998.
	
	\bibitem{kalman1960contributions}
	R.~E. Kalman.
	\newblock Contributions to the theory of optimal control.
	\newblock {\em Bol. Soc. Mat. Mexicana}, 5(2):102--119, 1960.
	
	\bibitem{kalman1960general}
	R.~E. Kalman.
	\newblock On the general theory of control systems.
	\newblock In {\em Proc. 1st IFAC World Congress, Moscow}, pages 481--492, 1960.
	
	\bibitem{karin}
	O.~Karin, A.~Swisa, B.~Glaser, Y.~Dor, and U.~Alon.
	\newblock Dynamical compensation in physiological circuits.
	\newblock {\em Mol. Syst. Biol.}, 12(11):886, 2016.
	
	\bibitem{anguelova2012efficient}
	J.~Karlsson, M.~Anguelova, and M.~Jirstrand.
	\newblock An efficient method for structural identiability analysis of large
	dynamic systems.
	\newblock In {\em 16th IFAC Symposium on System Identification}, volume~16,
	pages 941--946, 2012.
	
	\bibitem{kostyukovskii1968simple}
	Y.~M. Kostyukovskii.
	\newblock Simple conditions of observability of nonlinear controlled systems.
	\newblock {\em Avtomat. Telemekh}, (10):32--41, 1968.
	
	\bibitem{kou1973observability}
	S.~R. Kou, D.~L. Elliott, and T.-J. Tarn.
	\newblock Observability of nonlinear systems.
	\newblock {\em Information and Control}, 22:89--99, 1973.
	
	\bibitem{ligon2018genssi}
	T.~S. Ligon, F.~Fr\"ohlich, O.~T. Chi{\c{s}}, J.~R. Banga, E.~Balsa-Canto, and
	J.~Hasenauer.
	\newblock Genssi 2.0: multi-experiment structural identifiability analysis of
	sbml models.
	\newblock {\em Bioinformatics}, btx735, 2017.
	
	\bibitem{ljung1999system}
	L.~Ljung.
	\newblock {\em System identification: theory for the user}.
	\newblock Prentice Hall, Upper Saddle River, NJ, USA, 1999.
	
	\bibitem{ljung1994global}
	L.~Ljung and T.~Glad.
	\newblock On global identifiability for arbitrary model parametrizations.
	\newblock {\em Automatica}, 30(2):265--276, 1994.
	
	\bibitem{martinelli2015extension}
	A.~Martinelli.
	\newblock Extension of the observability rank condition to nonlinear systems
	driven by unknown inputs.
	\newblock In {\em Proceedings of the 23th Mediterranean Conference on Control
		and Automation (MED)}, pages 589--595. IEEE, 2015.
	
	\bibitem{martinelli2018nonlinear}
	A.~Martinelli.
	\newblock Nonlinear unknown input observability: Extension of the observability
	rank condition.
	\newblock {\em IEEE Trans. Autom. Control}, 2018.
	
	\bibitem{merkt2015higher}
	B.~Merkt, J.~Timmer, and D.~Kaschek.
	\newblock Higher-order lie symmetries in identifiability and predictability
	analysis of dynamic models.
	\newblock {\em Physical Review E}, 92(1):012920, 2015.
	
	\bibitem{meshkat2011finding}
	N.~Meshkat, C.~Anderson, and J.~J. DiStefano.
	\newblock Finding identifiable parameter combinations in nonlinear ode models
	and the rational reparameterization of their input--output equations.
	\newblock {\em Math. Biosci.}, 233(1):19--31, 2011.
	
	\bibitem{meshkat2009algorithm}
	N.~Meshkat, M.~Eisenberg, and J.~J. DiStefano.
	\newblock An algorithm for finding globally identifiable parameter combinations
	of nonlinear ode models using gr{\"o}bner bases.
	\newblock {\em Math. Biosci.}, 222(2):61--72, 2009.
	
	\bibitem{meshkat2014finding}
	N.~Meshkat, C.~E.-z. Kuo, and J.~DiStefano~III.
	\newblock On finding and using identifiable parameter combinations in nonlinear
	dynamic systems biology models and combos: A novel web implementation.
	\newblock {\em PLoS One}, 9(10), 2014.
	
	\bibitem{miao2011identifiability}
	H.~Miao, X.~Xia, A.~S. Perelson, and H.~Wu.
	\newblock On identifiability of nonlinear ode models and applications in viral
	dynamics.
	\newblock {\em SIAM Rev.}, 53(1):3--39, 2011.
	
	\bibitem{middendorf2016structural}
	T.~R. Middendorf and R.~W. Aldrich.
	\newblock Structural identifiability of equilibrium ligand-binding parameters.
	\newblock {\em J. Gen. Physiol.}, pages jgp--201611702, 2016.
	
	\bibitem{munoz2018or}
	R.~Mu{\~n}oz-Tamayo, L.~Puillet, J.-B. Daniel, D.~Sauvant, O.~Martin,
	M.~Taghipoor, and P.~Blavy.
	\newblock To be or not to be an identifiable model. {I}s this a relevant
	question in animal science modelling?
	\newblock {\em animal}, 12(4):701--712, 2018.
	
	\bibitem{Pohjanpalo1978}
	H.~Pohjanpalo.
	\newblock System identifiability based on the power series expansion of the
	solution.
	\newblock {\em Math. Biosci.}, 41(1):21--33, 1978.
	
	\bibitem{prajna2002introducing}
	S.~Prajna, A.~Papachristodoulou, and P.~A. Parrilo.
	\newblock Introducing sostools: A general purpose sum of squares programming
	solver.
	\newblock In {\em Proceedings of the 41st IEEE Conference on Decision and
		Control}, volume~1, pages 741--746. IEEE, 2002.
	
	\bibitem{procopio2017model}
	A.~Procopio, S.~De~Rosa, C.~Covello, A.~Merola, J.~Sabatino, A.~De~Luca,
	C.~Indolfi, F.~Amato, and C.~Cosentino.
	\newblock A model of cardiac troponin t release in patient with acute
	myocardial infarction.
	\newblock In {\em Proceedings of the 56th IEEE Conference on Decision and
		Control}, pages 435--440. IEEE, 2017.
	
	\bibitem{raman2017delineating}
	D.~V. Raman, J.~Anderson, and A.~Papachristodoulou.
	\newblock Delineating parameter unidentifiabilities in complex models.
	\newblock {\em Phys. Rev. E}, 95(3):032314, 2017.
	
	\bibitem{raue2014comparison}
	A.~Raue, J.~Karlsson, M.~P. Saccomani, M.~Jirstrand, and J.~Timmer.
	\newblock Comparison of approaches for parameter identifiability analysis of
	biological systems.
	\newblock {\em Bioinformatics}, page btt006, 2014.
	
	\bibitem{raue2009structural}
	A.~Raue, C.~Kreutz, T.~Maiwald, J.~Bachmann, M.~Schilling, U.~Klingm{\"u}ller,
	and J.~Timmer.
	\newblock Structural and practical identifiability analysis of partially
	observed dynamical models by exploiting the profile likelihood.
	\newblock {\em Bioinformatics}, 25(15):1923--1929, 2009.
	
	\bibitem{saccomani2003parameter}
	M.~P. Saccomani, S.~Audoly, and L.~D'Angi{\`o}.
	\newblock Parameter identifiability of nonlinear systems: the role of initial
	conditions.
	\newblock {\em Automatica}, 39(4):619--632, 2003.
	
	\bibitem{saccomani2018union}
	M.~P. Saccomani and K.~Thomaseth.
	\newblock The union between structural and practical identifiability makes
	strength in reducing oncological model complexity: A case study.
	\newblock {\em Complexity}, 2018, 2018.
	
	\bibitem{sedoglavic2002probabilistic}
	A.~Sedoglavic.
	\newblock A probabilistic algorithm to test local algebraic observability in
	polynomial time.
	\newblock {\em J. Symb. Comput.}, 33:735--755, 2002.
	
	\bibitem{sontag1982mathematical}
	E.~D. Sontag.
	\newblock {\em Mathematical control theory: deterministic finite dimensional
		systems}, volume~6.
	\newblock Springer Science \& Business Media, 2013.
	
	\bibitem{sontag2017dynamic}
	E.~D. Sontag.
	\newblock Dynamic compensation, parameter identifiability, and equivariances.
	\newblock {\em PLoS computational biology}, 13(4):e1005447, 2017.
	
	\bibitem{stigter2017assessing}
	J.~Stigter, M.~Beck, and J.~Molenaar.
	\newblock Assessing local structural identifiability for environmental models.
	\newblock {\em Environ. Model. Softw.}, 93:398--408, 2017.
	
	\bibitem{stigter2015fast}
	J.~D. Stigter and J.~Molenaar.
	\newblock A fast algorithm to assess local structural identifiability.
	\newblock {\em Automatica}, 58:118--124, 2015.
	
	\bibitem{sussmann1972controllability}
	H.~J. Sussmann and V.~Jurdjevic.
	\newblock Controllability of nonlinear systems.
	\newblock {\em J. Differ. Equations}, 12(1):95--116, 1972.
	
	\bibitem{tonsing2014cause}
	C.~T{\"o}nsing, J.~Timmer, and C.~Kreutz.
	\newblock Cause and cure of sloppiness in ordinary differential equation
	models.
	\newblock {\em Phys. Rev. E}, 90(2):023303, 2014.
	
	\bibitem{transtrum2015perspective}
	M.~K. Transtrum, B.~B. Machta, K.~S. Brown, B.~C. Daniels, C.~R. Myers, and
	J.~P. Sethna.
	\newblock Perspective: Sloppiness and emergent theories in physics, biology,
	and beyond.
	\newblock {\em J. Chem. Phys.}, 143(1):07B201\_1, 2015.
	
	\bibitem{tunali1987new}
	E.~T. Tunali and T.-J. Tarn.
	\newblock New results for identifiability of nonlinear systems.
	\newblock {\em IEEE Trans. Autom. Control}, 32(2):146--154, 1987.
	
	\bibitem{tuncer2018structural}
	N.~Tuncer, M.~Marctheva, B.~LaBarre, and S.~Payoute.
	\newblock Structural and practical identifiability analysis of zika
	epidemiological models.
	\newblock {\em Bull. Math. Biol.}, pages 1--33, 2018.
	
	\bibitem{vajda1989similarity}
	S.~Vajda, K.~R. Godfrey, and H.~Rabitz.
	\newblock Similarity transformation approach to identifiability analysis of
	nonlinear compartmental models.
	\newblock {\em Math. Biosci.}, 93(2):217--248, 1989.
	
	\bibitem{vidyasagar1993nonlinear}
	M.~Vidyasagar.
	\newblock {\em Nonlinear systems analysis}.
	\newblock Prentice Hall, Englewood Cliffs, NJ, 1993.
	
	\bibitem{villaverde2017dynamical}
	A.~F. Villaverde and J.~R. Banga.
	\newblock Dynamical compensation and structural identifiability of biological
	models: Analysis, implications, and reconciliation.
	\newblock {\em PLoS Comput. Biol.}, 13(11):e1005878, 2017.
	
	\bibitem{villaverde2017dynamicalarxiv}
	A.~F. Villaverde and J.~R. Banga.
	\newblock Dynamical compensation in biological systems as a particular case of
	structural non-identifiability.
	\newblock {\em arXiv preprint arXiv:1701.02562}, 2017.
	
	\bibitem{villaverde2017structural}
	A.~F. Villaverde and J.~R. Banga.
	\newblock Structural properties of dynamic systems biology models:
	Identifiability, reachability, and initial conditions.
	\newblock {\em Processes}, 5(2):29, 2017.
	
	\bibitem{villaverde2016identifiability}
	A.~F. Villaverde and A.~Barreiro.
	\newblock Identifiability of large nonlinear biochemical networks.
	\newblock {\em MATCH Commun. Math. Comput. Chem.}, 76(2):259--276, 2016.
	
	\bibitem{villaverde2016structural}
	A.~F. Villaverde, A.~Barreiro, and A.~Papachristodoulou.
	\newblock Structural identifiability of dynamic systems biology models.
	\newblock {\em PLoS Comput. Biol.}, 12(10):e1005153, 2016.
	
	\bibitem{villaverde2018input}
	A.~F. Villaverde, N.~D. Evans, M.~J. Chappell, and J.~R. Banga.
	\newblock Input-dependent structural identifiability of nonlinear systems.
	\newblock {\em IEEE Control Syst. Letters}, 3(2):272--277, 2019.
	
	\bibitem{walch2016parameter}
	O.~J. Walch and M.~C. Eisenberg.
	\newblock Parameter identifiability and identifiable combinations in
	generalized {H}odgkin--{H}uxley models.
	\newblock {\em Neurocomputing}, 199:137--143, 2016.
	
	\bibitem{walter1981unidentifiable}
	E.~Walter and Y.~Lecourtier.
	\newblock Unidentifiable compartmental models: what to do?
	\newblock {\em Math. Biosci.}, 56(1):1--25, 1981.
	
	\bibitem{walter1982global}
	E.~Walter and Y.~Lecourtier.
	\newblock Global approaches to identifiability testing for linear and nonlinear
	state space models.
	\newblock {\em Math. Comput. Simul}, 24(6):472--482, 1982.
	
	\bibitem{walter1997identification}
	E.~Walter and L.~Pronzato.
	\newblock {\em Identification of parametric models from experimental data}.
	\newblock Communications and Control Engineering Series. Springer, London, UK,
	1997.
	
	\bibitem{xia2003identifiability}
	X.~Xia and C.~H. Moog.
	\newblock Identifiability of nonlinear systems with application to {HIV/AIDS}
	models.
	\newblock {\em IEEE Trans. Autom. Control}, 48(2):330--336, 2003.
	
	\bibitem{yates2009structural}
	J.~W. Yates, N.~D. Evans, and M.~J. Chappell.
	\newblock Structural identifiability analysis via symmetries of differential
	equations.
	\newblock {\em Automatica}, 45(11):2585--2591, 2009.
	
\end{thebibliography}
\end{document}